# Interplay of Electrostatic Interaction and Steric Repulsion between Bacteria and Gold Surface Influences Raman Enhancement

*Jia Dong,[a] Jeong Hee Kim,[a] Isaac Pincus,[a] Sujan Manna,[a] Jennifer M. Podgorski,[d] Yanmin Zhu,[a] and Loza F. Tadesse [a, b, c]\**

a. Department of Mechanical Engineering, Massachusetts Institute of Technology, Cambridge, Massachusetts 02139, United States
b. Ragon Institute of MGH, MIT and Harvard, Cambridge, Massachusetts 02139, United States
c. Jameel Clinic for AI & Healthcare, Massachusetts Institute of Technology, Cambridge, Massachusetts 02139, United States
d. Cryo-EM Facility at MIT.nano, Department of Biology, Massachusetts Institute of Technology, Cambridge, Massachusetts 02139, United States

Corresponding Author:
\*Loza F. Tadesse, E-mail: lozat@mit.edu

**Abstract:** Plasmonic nanostructures have wide applications in photonics including pathogen detection and diagnosis via Surface-Enhanced Raman Spectroscopy (SERS). Despite major role plasmonics play in signal enhancement, electrostatics in SERS is yet to be fully understood and harnessed. Here, we perform a systematic study of electrostatic interactions between 785 nm resonant gold nanorods designed to harbor zeta potentials of +29, +16, 0 and -9 mV spanning positive neutral and negative domains. SERS activity is tested on representative Gram-negative *Escherichia coli* and Gram-positive *Staphylococcus epidermidis* bacteria with zeta potentials of -30 and -23 mV respectively in water. Raman spectroscopy and Cryo-Electron microscopy reveal that +29, +16, 0 and -9 mV nanorods give SERS enhancement of 7.2X, 3.6X, 4.2X, 1.3X to *Staphylococcus epidermidis* and 3.9X, 2.8X, 2.9X, 1.1X to *Escherichia coli*. Theoretical results show that electrostatics play the major role among all interaction forces in determining cell-nanorod proximity and signal enhancement. We identify steric repulsion due to cell protrusions to be the critical opposing force. Finally, a design principle is proposed to estimate the electrostatic strength in SERS. Our work provides new insights into the principle of bacteria-nanorod interactions, enabling reproducible and precise biomolecular readouts, critical for next-generation point-of-care diagnostics and smart healthcare applications.

# 1. Introduction

Surface-enhanced Raman spectroscopy (SERS) is a powerful technique that is widely used for live cell monitoring as well as detection of bacteria and biomolecular species.[1-3] To date, the modern tools used for infection diagnosis, such as polymerase chain reaction (PCR),[4] mass spectrometry[5] and lateral flow enzyme immunoassays,[6-7] are slow and destructive to cells. There is an emergent need to find an alternative method that is able to detect bacteria in a fast and accurate manner. SERS is a great candidate to achieve this purpose because it can rapidly detect pathogens even at a single-cell-level[8-10] while maintaining viability and versatility for both solid and liquid samples.[11-20] Gold nanorods are among the main plasmonic materials of choice to enhance the Raman signal of bacteria because of their robust synthesis protocol,[21] high stability and biocompatibility[22] making them suitable for clinical translation and biomedical applications[23-24] including cancer therapy,[25-27] drug testing[28-29] and pathogen detection.[30-31]

Clear understanding and systematic control over interaction between gold nanoparticles and bacteria is critical for achieving reliable and reproducible SERS.[32-46] In the past decades, numerous studies have focused on using surface modification and chemisorption approaches to design cell-gold nanoparticle interactions for disease treatment and healthcare diagnostics.[47-62] However, studies focusing on the basic principles of electrostatic interactions as well as other intermolecular forces are limited,[63] since gold nanoparticles can damage bacteria[64-65] and electrostatic interactions are labile and may be easily disrupted by ionic strength and pH.[63] It is known that the electrostatic cell-nanoparticles interaction is mostly mediated by teichoic acids in Gram-positive bacteria and lipopolysaccharides and phospholipids in Gram-negative bacteria.[66-67] Previous studies have also demonstrated the relationship between the zeta potential of different bacteria and the number of interacted gold nanoparticles.[68] Moreover, in our previous work, we have shown that in liquid, electrostatic interactions between nanoparticles and cells, in addition to the plasmonic effects, may have a significant role in determining the intensity of signal enhancement with up to 2X intensity difference.[17] However, detailed investigation of effects of the electrostatics and competition of other intermolecular forces is lacking, which renders this potentially powerful knob for tuning nanoparticle distribution towards reproducible SERS yet to be tapped.

Here, we perform a comprehensive study of local interactions between positive, neutral and negatively charged gold nanorods (+29, +16, 0 and -9 mV) and representative Gram-negative *Escherichia coli (E. coli)* and Gram-positive *Staphylococcus epidermidis (S.*

*epidermidis*) bacteria with surface charge of -30 and -23 mV respectively suspended in water as shown in Figure 1. Surface charges cannot go further below -9 mV due to the limit of the ligand exchange. Surface charge of nanorods is optimized by ligand exchange of the original positively charged surfactant cetyl trimethyl ammonium bromide (CTAB) with negatively charged sodium dodecyl sulfate (SDS), which maintains all other synthesis parameters consistent. SERS signatures for the different cell-nanorod mixtures show a clear relationship between interaction profiles of plasmonic nanorods and bacteria and their respective signal enhancement profiles. Each sample mixture was tested 120 times in four repeated experiments to ensure the reproducibility of the measurements and the trend we obtained. Based on both our experimental results and theoretical calculations, we found that electrostatic interactions play the most critical role among all molecular interaction forces including steric repulsions in determining SERS activities. Compared to Raman intensity of cells alone, we observed signal enhancements leading to 7.2 ± 0.7X, 3.6 ± 0.3X, 4.2 ± 0.2X, 1.3 ± 0.0X for S. *epidermidis* and 3.9 ± 0.2X, 2.8 ± 0.0X, 2.9 ± 0.2X, 1.1 ± 0.0X for *E. coli* mixed with +29, +16, 0 and -9 mV nanorods, respectively. Particularly, the higher signal enhancement of 0 mV nanorod compared to 16 mV nanorod comes from a larger number of hotspots due to heavy agglomeration. This finding rigorously illuminated the effects of cell-nanorod interaction on SERS activities, confirming that electrostatic interaction plays a major role in controlling SERS enhancement, which will be valuable to achieve a reproducible and controllable SERS towards successful clinical transition and various smart healthcare applications.

## 2. Results and Discussion

The SERS measurements were conducted using a liquid sample well previously described (Figure 1a).[17] High contrast Differential Interference Contrast (DIC) images confirmed the rod shape of *E. coli* and round shape of *S. epidermidis* (Figure 1b). The lengths of *E. coli* and *S.epidermidis* used in our study are approximately 2 and 0.5 μm, measured by our bright field microscopy (Figure S1). Zeta potential measurements of bacteria[69] obtained from the average of three biological replicates show *E. coli* and *S. epidermidis* in water have -30 and -23 mV surface charge respectively (Figure S2). The gold nanorods used in this study were synthesized by a seed growth method.[70] The seed solution was first prepared by adding $NaBH_4$ into $HAuCl_4$ to reduce the gold precursor. Then, $AgNO_3$, ascorbic acid, CTAB and gold precursors were mixed with the gold seed solution to grow the gold nanorod into the desired aspect ratio (48 ± 5 and 14 ± 2 nm). The positively charged CTAB gives the as-synthesized gold nanorods a zeta potential of 29 mV (Figure S2).

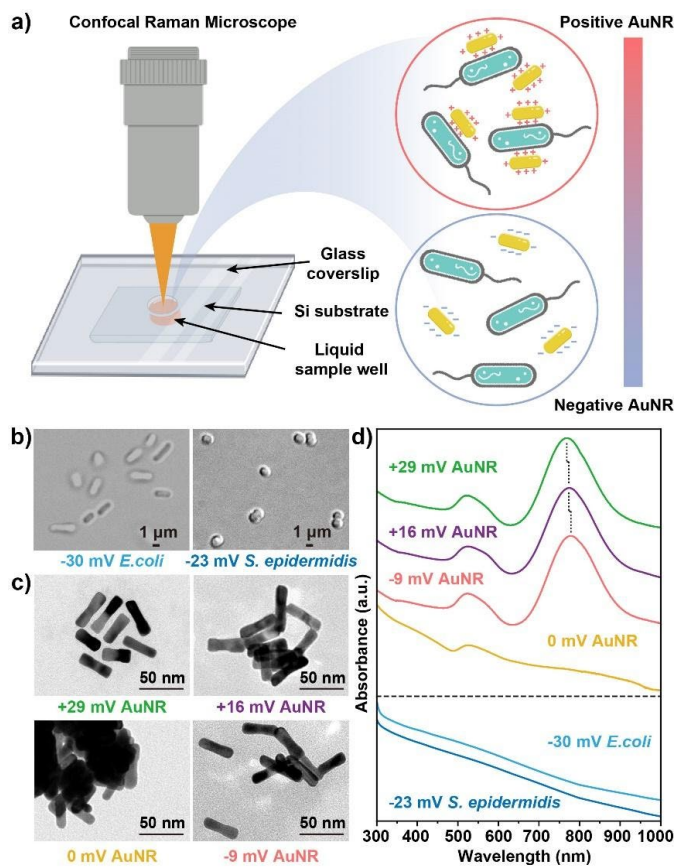

**Figure 1.** Experimental setup and materials used in this study. (a) Schematic showing the setup of a liquid sample well, which is composed of silicon substrate and glass coverslip connected by double-sided tape. The bacteria mixed with gold nanorods (AuNR) of different zeta potentials were placed in the central hole of the liquid sample well. (b) Differential interference contrast (DIC) images of rod-shaped *E. coli* (left) and round *S. epidermidis* (right) used for liquid SERS measurement in this study. (c) Transmission electron microscopy (TEM) images of +29, +16, 0 and -9 mV gold nanorods used for liquid SERS measurement, showing rod shape and size is maintained after surfactant exchange is performed. Notably 0 mV is aggregated as there is no inter-particle repulsion. (d) UV-Vis-NIR spectra of +29, +16, 0 and -9 mV gold nanorods and bacteria. The spectra of 0 mV gold nanorods does not show any notable absorption peak due to heavy agglomeration, while slight red shifts are observed for nanorod plasmonic peaks when zeta potential decreases from +29 to +16 and then to -9 mV.

Surface charge of gold nanorods was adjusted by surfactant exchange with negatively charged sodium dodecyl sulfate (SDS) by varying reaction time and SDS concentration. As a result, gold nanorods with zeta potential of +16, 0 and -9 mV were obtained (Figure S2). Transmission electron microscopy (TEM) of gold nanorod samples shows no obvious changes of aspect ratios among all four +29, +16, 0 and -9 mV gold nanorods (Figures 1c and S3).

However, we did observe some agglomeration of +16, 0 and -9 mV gold nanorods compared to the original +29 mV gold nanorods after the ligand exchange process of CTAB with SDS. The 0 mV gold nanorods have the most agglomeration, and rapid precipitation out of solution (Figure. S4). UV-Vis-NIR absorption spectra for +29, +16 and -9 mV gold nanorods show two absorption peaks at 510 and 770 nm, which are assigned to the transverse localized surface plasmon resonance and longitudinal localized surface plasmon resonance bands respectively (Figure 1d). The longitudinal plasmonic band is gradually red shifted (770 nm – 774 nm – 778 nm) as the zeta potential drops from +29 to +16 and then to -9 mV probably due to slight size increment after CTAB exchange with SDS. The absorption spectra measurement for 0 mV gold nanorods does not show a distinct peak, which is a typical feature of large aggregation and lack of monodispersity in solution. We also measured the absorption spectra for pure *S. epidermidis* and *E. coli*, which show no distinct absorption peaks.

We mixed the bacteria and gold nanorod with a 1:1 volumetric ratio for SERS measurements as described previously.[17] Before mixing, both bacteria and gold nanorods were washed with water to decrease the side effects of protecting ligands or excess growth media. We conducted the SERS measurements using a commercial WiTec Raman microscope equipped with a 785 nm laser. Each spectrum was obtained from an average of 30 measurements. SERS spectra of *S. epidermidis* and *E. coli* mixed with different zeta potentials of gold nanorods are provided in Figure 2a and b. All samples show strong SERS signatures, giving 7.2 ± 0.7X, 3.6 ± 0.3X, 4.2 ± 0.2X, 1.3 ± 0.0X for S. *epidermidis* and 3.9 ± 0.2X, 2.8 ± 0.0X, 2.9 ± 0.2X, 1.1 ± 0.0X for *E. coli* mixed with 29, 16, 0 and -9 mV nanorods, respectively. While greater SERS amplification, reaching several orders of magnitude, has been reported when using a high numerical aperture of objective lens in a targeted, controlled condition, we would like to note that our SERS studies were conducted in a liquid environment with dynamic movements of cells and nanorods and low numerical aperture of objective lens. The design of sealed liquid wells also makes it difficult for light to focus on the middle of the liquid droplet and pass through the coverslip. All the above factors result in fewer Raman scatter light reaching the detector different from the traditional SERS measurements normally giving larger orders of magnitude enhancement. Among all nanorods, the one with +29 mV gives the highest signal enhancement, followed by the others in an order of +29 mV > 0 mV > +16 mV > -9 mV. This observation is further corroborated by plotting the area under the Raman bands for signature bacterial peaks near 1000, 1300, 1400 and 1500 cm$^{-1}$ (Figure 2c and d). Starting from the highest SERS signal given by +29 mV nanorods, the SERS enhancement obviously decreases as the zeta potential of gold nanorods drops from

+29 to +16 and then to -9 mV. However, we didn't observe any obvious signal drops from +16 to 0 mV, indicating that the gold nanorod with 0 mV zeta potential is unique compared to other positive and negative nanorods because of its large aggregation leading to larger numbers of hotspots, thus enhancement is comparable to +16 mV nanorods due to close cell-nanorod contact. Next, we used heatmap to check the consistency of enhancements across the scanning area for different cell-nanorod combinations (Figure 2a and b). As expected, +29, +16, 0 and -9 mV nanorods all show consistent signal enhancement across the scanned area (Figure S5). Thus, our SERS experiment is highly reproducible, and all zeta potentials of nanorods are able to provide uniform Raman signal enhancement.

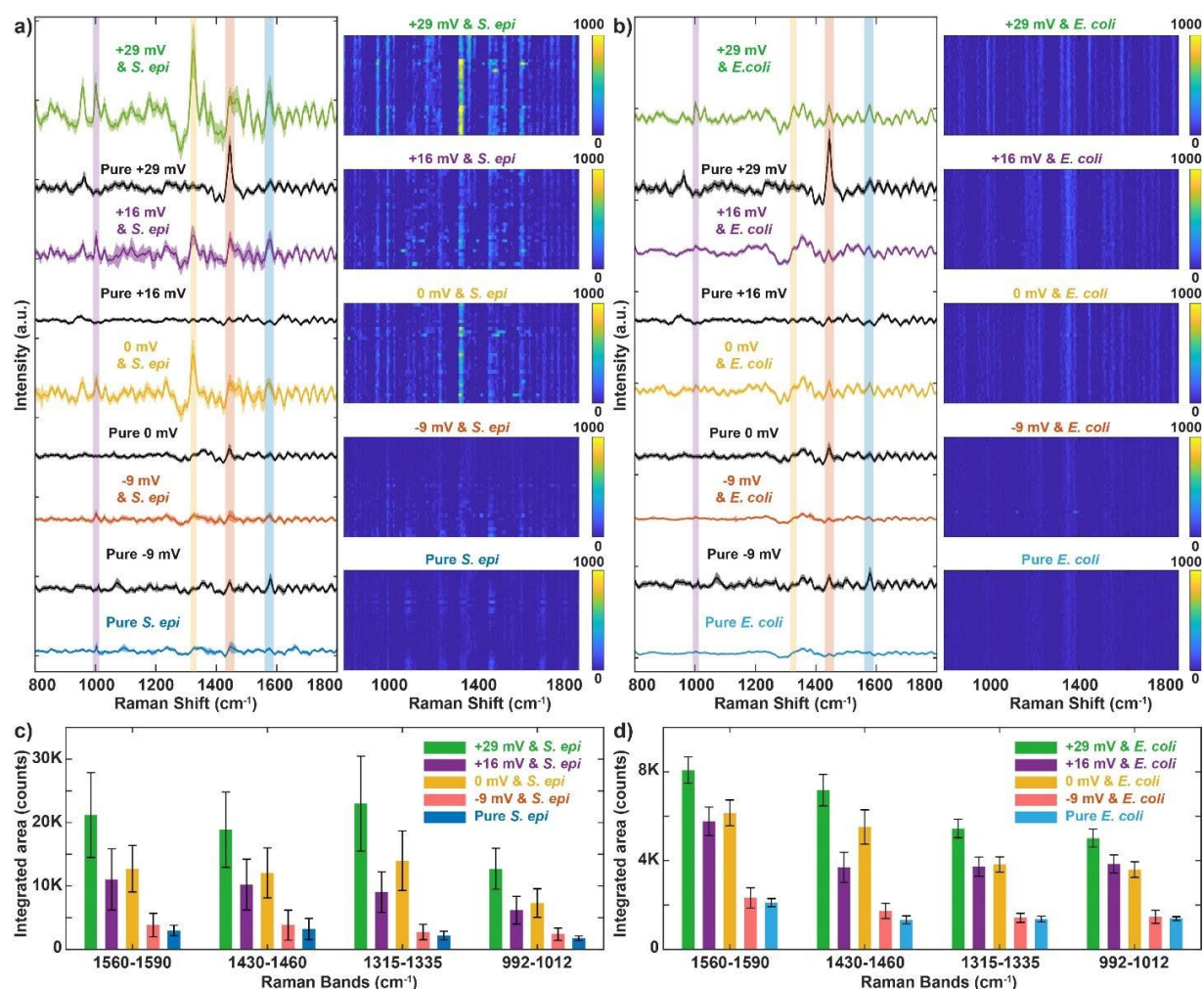

**Figure 2.** SERS spectra comparison for different cell-nanorod mixtures with standard deviations. (a and b) Mean SERS signatures of *E. coli* and *S. epidermidis* mixed with +29, +16, 0 and -9 mV nanorods, suggesting that *S. epidermidis* generally have higher signal intensities than *E. coli*. The heatmap plots are also shown next to Raman spectra. The yellow or bright regions show higher signal intensity and the uniform SERS signatures indicate the consistency of enhancements within different scanning areas and repeats. The Raman spectra of pure nanorods and bacteria are also provided for comparison. (c and d) Bar plots showing

integrated areas for representative bacterial Raman bands near 1000, 1300, 1400 and 1500 cm$^{-1}$. The SERS activity decreases together with the drops of nanorod zeta potentials in an order of +29 mV > 0 mV > +16 mV > -9 mV. The plotted spectra are average and standard deviation of 30 spectra from each sample well. Additional three biological repeats with 90 spectral data points are shown in Figure S6-9.

When we compare the SERS signatures between *E. coli* and *S. epidermidis*, we found that for same zeta potential of nanorod, *S. epidermidis* always has higher intensity spectra than *E.coli* consistent with our previous report (Figure 2c and d).[17] When the +29 mV nanorod was used, nearly 3.3X difference of signal enhancement was observed between the two bacteria (7.2X for *S. epidermidis*, 3.9X for *E. coli*). As the zeta potential of nanorods decreases to +16 and 0 mV, the difference of spectra enhancement diminishes accordingly, changing to 0.8X and 1.3X respectively. The -9 mV nanorod only shows a tiny 0.2X difference of signal enhancement between *E. coli* and *S. epidermidis*. Our results clearly show that the type of bacteria could also affect electrostatic interactions and hence Raman signal enhancement profile.

In order to confirm the reproducibility of the results we obtained from the SERS measurements, we further performed three additional biological repeats generating 90 spectral data points measured on different dates for each cell-nanorod mixture (Figures S6 and S7). The same enhancement profile of +29 mV > 0 mV > +16 mV > -9 mV nanorod was obtained (Figure S8), strongly corroborating the results provided in Figure 2. In addition, uniform Raman signal enhancement was observed for all repeats similar to results that we demonstrated in Figure 2 showing the robust liquid well sample preparation protocol (Figure S9). We however observe that the reproduced spectra show some subtle signature differences compared to the original spectra. For example, the SERS signature peak of *E. coli* around 1000 cm$^{-1}$, which was assigned to the L-phenylalanine from metabolic activity,[71] is not prominent in spectra shown in Figure 2 but appears to be strong peak in additional repeats (Figures S6 and S7). This is due to the nature of SERS fluctuations that certain parts of bacteria signature can be selectively enhanced depending on proximity of molecules landing in hotspots.

Our Raman results show a strong relationship between SERS enhancement and surface charges of both the bacteria and nanorods. Since both *E. coli* and *S. epidermidis* have negative surface zeta potentials, as all bacteria do, positive gold nanorod generally gives higher signal enhancement due to the large electrostatic attraction. In contrast, negative gold nanorods show less signal enhancement for both bacteria due to large electrostatic repulsion.

To further visualize the local interaction between the various bacteria nanorod mixtures, we performed Cryogenic Electron Microscopy (Cryo-EM) imaging by freezing the cell-nanorod mixtures onto 200-mesh Cu lacey carbon grid, which can closely capture the interactions between the cells and nanorods as-is in liquid (Figure 3). As shown in Figure 3a, the +29 mV nanorods adhere closely and evenly to both of *S. epidermidis* and *E. coli* membranes. When the zeta potential of nanorods decreases to +16 mV (Figure 3b), nanorods start to aggregate while still sticking onto the bacteria surface, resulting in less surface coverage of the cells and lower Raman signal enhancement. As the zeta potential of nanorods further drops to 0 mV (Figure 3c), there are no obvious cell-nanorod interactions and nanorods change to randomly spread around the cells. Notably, in some cases we observe that cells can be surrounded by large aggregates of 0 mV nanorods. This is because the 0 mV nanorods prefer to aggregate with each other due to zero inter-particle repulsion and occasionally cells are wrapped inside (Figure S10). The larger number of hotspots induced by aggregation wrapping the cells leads to higher signal enhancement of 0 mV nanorods compared to +16 mV nanorods. In the case of -9 mV nanorods, there exists large repulsion between the cells and nanorods and very few nanorods are observed on the bacterial surface. As a result, -9 mV nanorods give the lowest signal enhancement to both *S. epidermidis* and *E. coli* among all types of nanorods. Our cryo-EM images thus strongly corroborate our trends observed from SERS spectra shown in Figure 2.

To clearly understand if electrostatic interactions play the primary role in SERS enhancement, we further performed theoretical calculations that provide insight into the local interaction forces between the cells and nanorods. To model the cell-nanoparticle interaction, we consider both objects as colloidal particles interacting through a combination of Derjaguin–Landau–Verwey–Overbeek (DLVO) forces and steric polymer repulsion (Figures 4a and S11). Detailed calculation codes are provided in Supplementary Information (Table S1). Specifically, we calculate the free energy $\Delta G$ with respect to nanoparticle-bacteria separation distance $d$, including the effects of four separate forces. The first two are the electrostatic double-layer (ES) and Van der Waals (VDW) forces as in DLVO theory.[72] Applying the Derjaguin approximation and considering two constant-potential spheres of surface-area equivalent radii $a_1$ and $a_2$ with potentials $\psi_1$ and $\psi_2$, the free energy is given by:[73-74]

$$\Delta G^{ES} = \pi \varepsilon \frac{a_1 a_2}{a_1 + a_2} (\psi_1^2 + \psi_2^2) \left[ \frac{2\psi_1 \psi_2}{\psi_1^2 + \psi_2^2} \ln \frac{1 + exp(-\kappa d)}{1 - exp(-\kappa d)} + \ln\{1 - exp(-2\kappa d)\} \right]$$

$$\Delta G^{VDW} = -\frac{A a_1 a_2}{6d(a_1 + a_2)}$$

where $\kappa$ is the inverse Debye length (here $\frac{1}{\kappa} = 961$ nm for pure water), $\varepsilon$ is the permeability of water ($6.9 \times 10^{-10}$ F/m), and $A$ is the Hamaker constant (approximated as $3 \times 10^{-20}$ J). The potentials $\psi_1$ and $\psi_2$ are taken as the measured zeta potentials of the bacteria and nanoparticles. Note that we have assumed constant-potential surfaces, which leads to a purely attractive electrostatic potential for $\kappa d \ll 1$, irrespective of the signs of $\psi_1$ and $\psi_2$.[72] On the other hand, if the surface charge is instead constant, the force will be similar at long distances but purely repulsive for $\kappa d \ll 1$.

We further include a short-range repulsive potential as per the extended DLVO theory, which considers acid-base (AB) interactions:[72, 74]

$$\Delta G^{AB} = 2\pi\lambda\Delta G_0^{AB} \frac{a_1 a_2}{a_1 + a_2} exp\,[(d_0 - d)/\lambda]$$

here $\lambda = 0.6$ nm is the correlation length of molecules in the liquid medium, $d_0 = 0.157$ nm is the length of the closest approach, while $\Delta G_0^{AB} = 10^{-2} J/m^2$ is the strength of interaction. While we do not know the exact form of this potential, there virtually always exists some form of highly repulsive short-range potential between surfaces, which is captured in our model by $\Delta G^{AB}$.

Lastly, we postulate that there exist additional steric repulsions (SP) arising from biopolymers on the bacterial surface, which oppose the (attractive) DLVO forces. Prior theoretical work has suggested several possible forms for this interaction, generally based upon a mix of reduction in entropy and direct compression upon interaction between two polymer-laden surfaces.[72, 75-77] We have used expressions reported in the literature [58] which were explicitly tested against force-distance curves for biopolymers extruded by *Pseudomonas putida*:[78]

$$\Delta G^{SP} \approx 50 k_B T L^2 \Gamma^{3/2} \frac{a_1 a_2}{a_1 + a_2} exp\,-2\pi d/L$$

where $L$ is the extended length of the polymer layer on the bacterial surface (200 nm for *E. coli* but only 10 nm for *S. epidermidis*), while $\Gamma = 3 \times 10^{14}\ 1/m^2$ is the number of polymers per area on the bacterial surface (this value corresponds to one surface-attached biopolymer every ~60 nm).

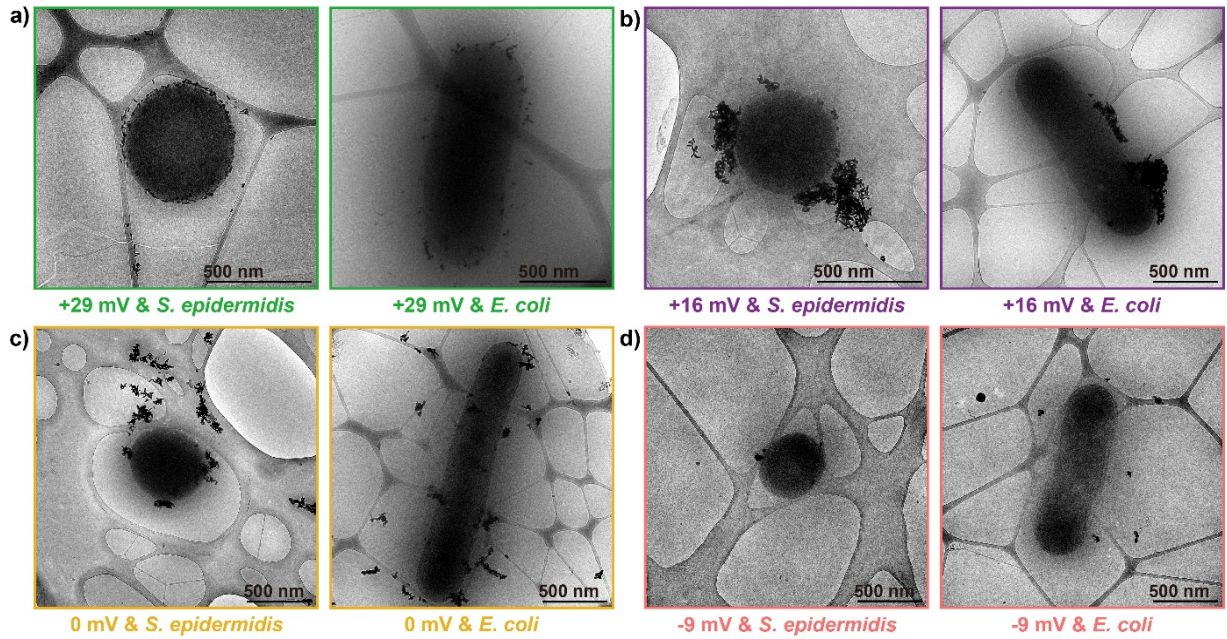

**Figure 3.** Cryo-electron microscopy images of local interaction between cells and nanorods. Cryo-EM images of *S.epidermidis* (left column) and *E. coli* (right column) interacting with (a) 29 mV, (b) 16 mV, (c) 0 mV and (d) -9 mV gold nanorods. Nanorods first surround bacteria well, aggregate with each other around the bacteria surface and are repelled away as zeta potential decreases.

Our modeling relies upon several assumptions, some of which will certainly affect quantitative accuracy. The Derjaguin approximation is strictly accurate only in the limit of $d \ll a$, or the range of the forces is much smaller than the particles.[79] We have assumed that the bacteria and particle surfaces remain at constant potential, the accuracy of which depends on details of charge regularization at the surfaces.[72] Further, the charges are assumed to lie on the surface of the bacteria, while in reality the extended biopolymers themselves are ionizable,[78] which affects the distance dependence of the double layer forces. Finally, the polymeric steric repulsions are derived through considering a straightforward balance of osmotic pressure and elastic energy, while more exact expressions can be found via self-consistent field theory.[80] The above approximations provide accurate estimations to cell-nanorod interactions while minimizing the computational cost.

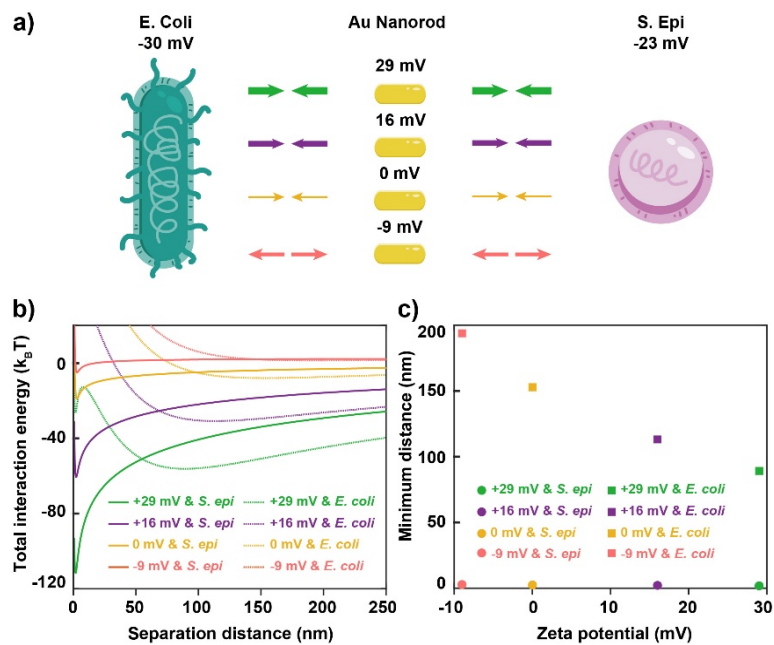

**Figure 4.** The cell-nanorod interactions investigated by theoretical calculations. (a) Schematic representation of cell-nanorod interactions between +29, +16, 0 and -9 mV gold nanorods and two bacteria (*E. coli* and *S. epidermidis*). (b) Theoretical calculations based on DLVO theory and steric polymer repulsion of different cell-nanorod interactions demonstrated in (a). The total interaction energy is modelled as a function of cell-nanorod separation distance *d*. The interaction energy approaches zero at a large distance as nanorod zeta potential decreases from +29 to -9 mV. (c) The plot of minimum separation distance with respect to zeta potential of nanorods. *S. epidermidis* has shorter equilibrium separation distances for all zeta potentials of nanorods compared to *E. coli*. Detailed values of minimum separation distance and corresponding interaction energies are provided in Table S2. Energy is reported in units of $k_BT$.

By plotting the interaction energy *ΔG* with respect to nanorod-cell separation *d*, we obtained potential energy profiles for cell-nanorod interactions, as shown in Figure 4b. The theoretical results show that at the same zeta potential of gold nanorods, *S. epidermidis* have closer distance to the nanorods, which further leads to almost 2X larger interaction energy than that of *E. coli* (Table S2). When the nanorod zeta potential decreases from +29 to -9 mV, the interaction energy drops accordingly and approaches zero at larger distances. This can be seen more clearly from Figure 4c when we plot the separation distance *d* with respect to the zeta potential of gold nanorods. For all nanorod zeta potentials, *S. epidermidis* has shorter equilibrium separation distances than *E. coli*. To better understand which type of force dominates the interaction energy *ΔG*, we plotted similar energy profiles for each force with respect to the separation distance *d* (Figure 5a and b). We found that within a short distance range (below 200 nm for *E. coli* and 10 nm for *S. epidermidis*), the interaction energy is mainly determined by both electrostatic interaction and steric repulsion. When the separation distance *d* grows beyond this range, electrostatic interaction solely becomes the dominant force. We also found that the rest of the two forces, i.e. acid-base interaction and Van der Waals force, also play roles in determining the final interaction energy *ΔG* throughout all the separation distance. However, their energies are small compared to electrostatic interaction and steric repulsion, giving a negligible contribution to the overall cell-nanorod interaction. In the case of 29 mV nanorod mixed with *S. epidermidis* (Figure 5a) and *E. coli* (Figure 5b), the combined effect of Van der Waals, electrostatic interaction, acid-base interaction and steric repulsion finally leads to minimum interaction energy of -56.2 and -111.6 $k_BT$ at separation distance of 89.2 and 1.8 nm for *E. coli* and *S. epidermidis* respectively (Figure 4b and c). Detailed quantitative results for other cell-nanorod mixtures are shown in Table S2.

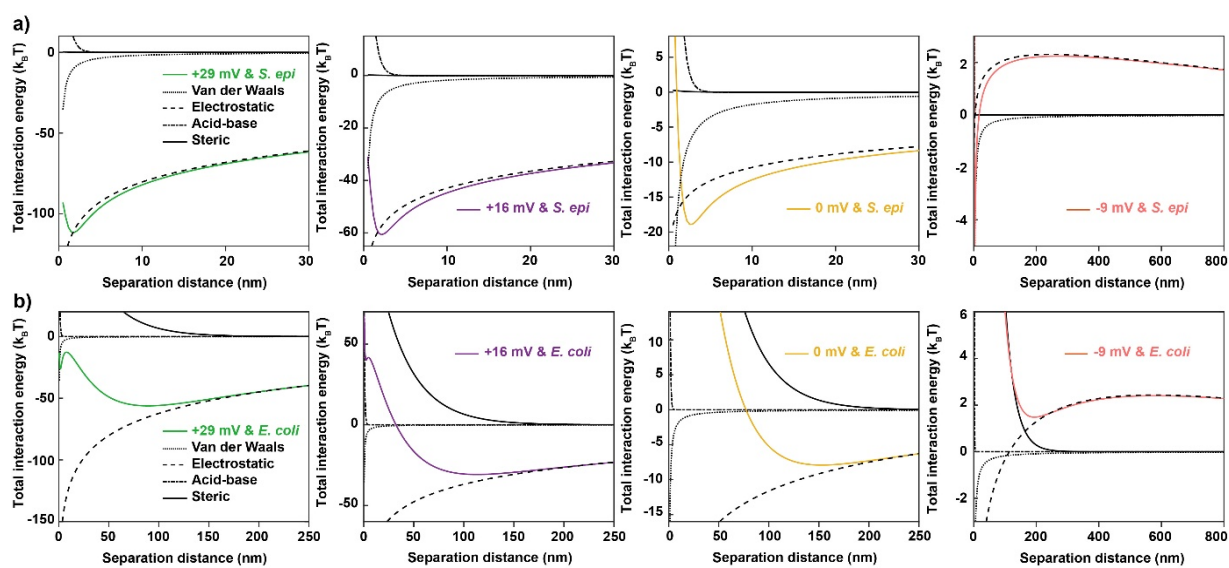

**Figure 5.** Split energy diagram of different cell-nanorod interactions investigated by theoretical calculations. (a) Detailed interaction energies between *S. epidermidis* and nanorods. (b) Detailed interaction energies between *E. coli* and nanorods. The total interaction energies are split into curves of single Van der Waals forces, electrostatic interactions, acid-base interactions and steric repulsions versus the separation distance *d*. Electrostatic interaction is the dominant force in all separation distances while steric repulsion becomes the critical opposing force in distance < 200 nm. Energy is reported in units of $k_BT$.

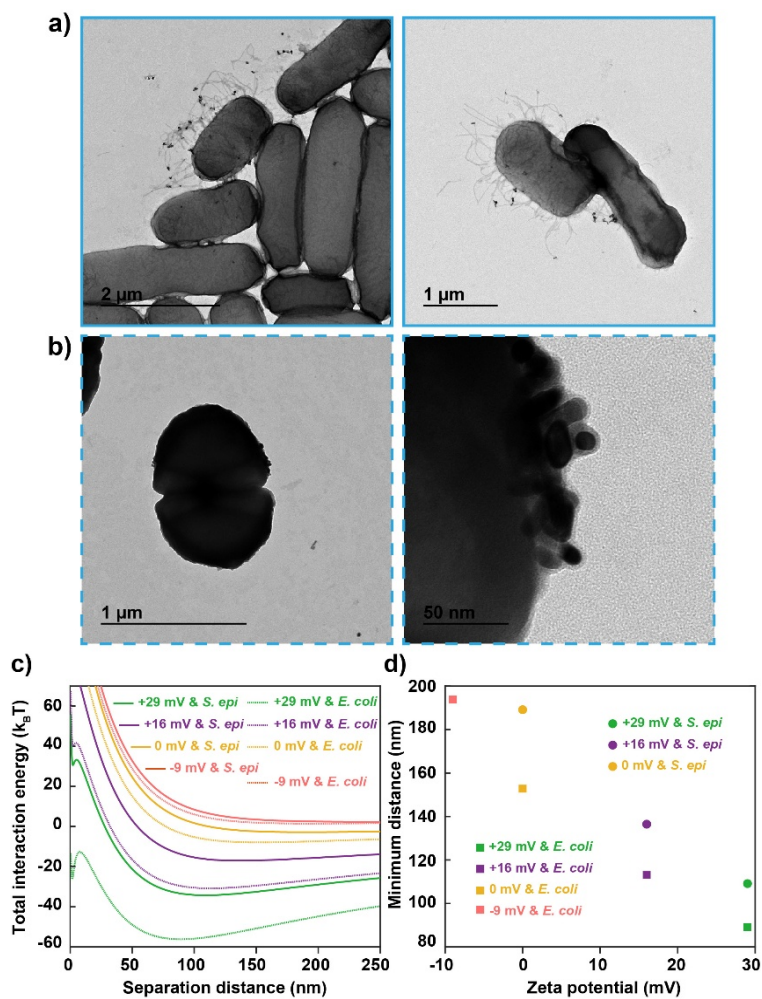

**Figure 6.** The influence of steric repulsion towards cell-nanorod interaction. Negative-stain TEM images of (a) *E. coli* and (b) *S. epidermidis* mixed with +29 mV gold nanorods showing that nanorods adhere tightly on *S. epidermidis* surface, whereas the bulky surface polymers of *E. coli* prevent nanorods from penetrating inside. (c) Plots of interaction energy versus the separation distance $d$ between cells and gold nanorods under the assumption that both *E. coli* and *S. epidermidis* have the same type of surface long-chain polymers. (d) Plots of minimum separation distance with respect to zeta potential of nanorods under the same assumption. Note that *S. epidermidis* and -9 mV nanorods do not have a minimum interaction energy. The new trend shown here is opposite to the results shown in Figures 4b and c, showing that steric repulsion could decrease the final interaction energy and increase minimum separation distance. Detailed values of minimum separation distance and corresponding interaction energies are provided in Table S3. Energy is reported in units of $k_BT$.

Interestingly we also discovered that steric repulsion could significantly influence the interaction energy $\Delta G$ within short range. Since *E. coli* has more negative zeta potential (-30 mV) compared to that of *S. epidermidis* (-23 mV), initially we expected that +29 mV nanorods could give more SERS enhancement to *E. coli* than *S. epidermidis*. Nevertheless, both Raman and Cryo-EM results have shown that *S. epidermidis* has better signal enhancement than *E. coli* as nanorods bind closely to its bacterial membrane surface. To resolve this further we performed negative-stain TEM images of dried samples of *S. epidermidis* and *E. coli* mixed with +29 mV gold nanorods. Importantly, we observed many long-chain polymers on *E. coli* membrane surface (Figure 6a), whereas similar polymers were not observed on *S. epidermidis* membrane surface (Figure 6b). Previous study[81] has shown that the long-chain polymers of *E. coli* are mainly composed of lipopolysaccharides, sugars and proteins that contribute to different cell surface properties such as zeta potential and hydrophobicity. Such polymers are so bulky that the nanorods could not easily penetrate inside to have close contact with the *E. coli* membrane surface. Although protrusions and metabolic wastes around the *E. coli* also contribute to Raman signatures, our results suggest that closer interaction with the membrane wall leads to higher signals. On the contrary, the lack of long-chain polymers in *S. epidermidis* enables nanorods to easily approach the membrane walls. Moreover, when nanorods are close to the membrane, they will be in close proximity to each other, creating dimeric hotspots that are stronger than further away located pairs of particles. All these differences result in stronger Raman signatures of *S.epidermidis* compared to *E. coli*. As a control experiment, we calculated the interaction energy *ΔG* with respect to nanorod-cell separation distance *d* (Table S3) by assuming both *S. epidermidis* and *E. coli* have the same type of surface long-chain polymers (Figures 5c and d) and observed an opposite trend compared to Figure 4b and c. This indicates that surface long-chain polymers of *E. coli* significantly influence its cell-nanorod interaction hence its overall SERS profile. All calculation results agree well with what we discovered from SERS experiments and Cryo-EM images, again confirming the importance of local cell-nanorod interactions for reproducible SERS.

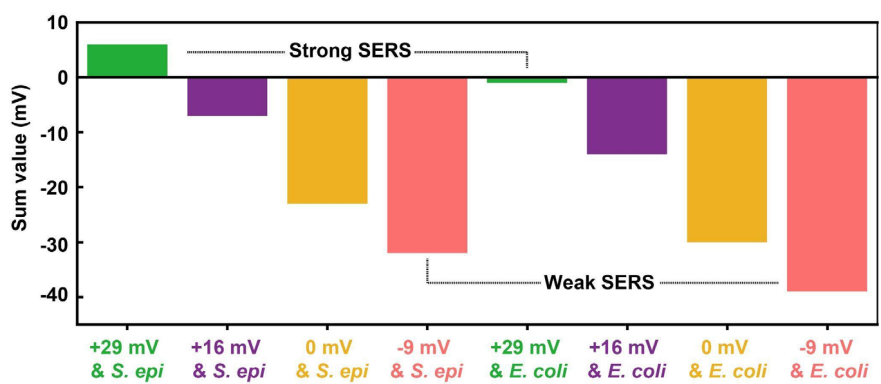

**Figure 7.** The zeta potential sum value chart of various cell-nanorod mixtures showing an easy approach to estimate the SERS activity.

Finally, based on our findings we would like to propose a straightforward approach to qualitatively determine expected SERS intensities via determination of strength of electrostatic interactions by simple sum of the zeta potential of bacteria and nanorods (Figure 7 and Table S2).

$$\Delta G^{ES} = \psi_{Bacteria} + \psi_{Nanorods}$$

This equation is much simpler than the standard electrostatic equation and can be used to make quick decisions about SERS intensities.[73] When the sum value is positive due to higher zeta potential of gold nanorods than bacteria, nanorods prefer to adhere closely to bacteria surface and a higher SERS enhancement can be expected. If gold nanorods have lower zeta potential than bacteria, on the contrary, the sum value would be negative, resulting in lower or even no SERS enhancement. In short, the prerequisite to see obvious SERS enhancement is to get a sum value at least higher than bacteria zeta potential itself. Any sum values lower than that will not lead to any obvious signal enhancement, as seen in the case of -9 mV nanorods mixed with *E. coli* and *S. epidermidis* in Figure 2.

## 3. Conclusion

Our work utilized both experiments and theoretical methods and revealed two key findings: (1) electrostatic interaction between bacteria and nanoparticles is a critical factor in determining cell-nanorod proximity and hence signal enhancement, (2) Cell protrusions, i.e. surface-associated biopolymers, create steric repulsions that become the key opposite forces to compete with electrostatic interactions. We demonstrated using two model bacteria, Gram-positive *E. coli* and Gram-negative *S. epidermidis*, and plasmonic gold nanorods with zeta potentials of +29, +16, 0 and -9 mV. The results show that +29, +16, 0 and -9 mV gold nanorods give around 7.2 ± 0.7X, 3.6 ± 0.3X, 4.2 ± 0.2X, 1.3 ± 0.0X signal enhancement for *S. epidermidis* and 3.9 ± 0.2X, 2.8 ± 0.0X, 2.9 ± 0.2X, 1.1 ± 0.0X signal enhancement for *E. coli* respectively. In addition, *S. epidermidis* consistently shows higher signal enhancements than *E. coli* when they interact with the same type of nanorod. Theoretical calculations further indicate that electrostatic interaction mainly determines the final SERS activity, whereas within a short distance below 200 nm the SERS activity is also influenced by steric repulsion depending on the length of polymers on bacteria surface. We also proposed a simple equation to roughly estimate the performance of signal enhancement for cell-nanorod mixtures. Our findings demonstrate the critical importance of electrostatic interaction for liquid bacterial SERS and provide principles for fabricating the designer plasmonic substrate towards reproducible SERS-based precision diagnostics, clinical translations and various smart healthcare applications.

**Acknowledgements**

Cryo-EM specimens were prepared and data was collected at the Cryo-EM Facility at MIT.nano, including use of the Talos Arctica gifted by the Arnold and Mabel Beckman Foundation. We would also like to thank Dr. DongSoo Yun (MIT David H. Koch Institute) for the help of Cryo-EM and Negative-Stain EM experiments and Dr. Jeffrey Kuhn (MIT David H. Koch Institute) for the help of DIC imaging.

**Conflict of Interest**

The authors declare no conflict of interest.

**Data Availability Statement**

The data that support the findings of this study are available from the corresponding author upon reasonable request.

Received: ((will be filled in by the editorial staff))
Revised: ((will be filled in by the editorial staff))
Published online: ((will be filled in by the editorial staff))

Supporting Information

# Interplay of Electrostatic Interaction and Steric Repulsion between Bacteria and Gold Surface Influences Raman Enhancement

*Jia Dong,[a] Jeong Hee Kim,[a] Isaac Pincus,[a] Sujan Manna,[a] Jenn M. Podgorski,[d] Yanmin Zhu,[a] and Loza F. Tadesse [a, b, c]\**

a. Department of Mechanical Engineering, Massachusetts Institute of Technology, Cambridge, Massachusetts 02139, United States
b. Ragon Institute of MGH, MIT and Harvard, Cambridge, Massachusetts 02139, United States
c. Jameel Clinic for AI & Healthcare, Massachusetts Institute of Technology, Cambridge, Massachusetts 02139, United States
d. Cryo-EM Facility at MIT.nano, Department of Biology, Massachusetts Institute of Technology, Cambridge, Massachusetts 02139, United States

Corresponding Author:
\*Loza F. Tadesse, E-mail: lozat@mit.edu

## 1. Materials

***Chemicals***: cetyltrimethylammonium bromide (CTAB, ≥98%, MP Biomedicals), Gold(III) chloride trihydrate (HAuCl$_4$·3H$_2$O, ≥99.9% trace metals basis, Sigma-Aldrich), sodium dodecyl sulfate (SDS, ≥99.0%, Sigma-Aldrich), sodium borohydride (NaBH$_4$, ≥98.0%, Sigma-Aldrich), silver nitrate (crystals, extra pure, Sigma-Aldrich), L-ascorbic acid (≥99.0%, VMR chemicals), deinoized water (DI water, Certified ACS, Fisher Scientific), acetone (≥99.5%, Certified ACS, Fisher Scientific). All chemicals were directly used as received.

***Bacteria Cells***: *E. coli* strain Seattle 1946 cells and *S. epidermidis* FDA strain PCI 1200 (ATCC #25922 and #12228, respectively) were used in this study.

## 2. Experimental Methods

***Synthesis of original CTAB-stabilized gold nanorods (zeta potential: 29 mV)***: The synthesis of the original gold nanorods with positively charged surfactant CTAB was conducted according to literature procedures.[1] It is worth noting that the sodium oleate was not used during the synthesis, so CTAB is the only surfactant protecting the nanorod surface. The as-synthesized gold nanorods were collected by centrifuging 30 mL aliquots and further washed with 1 mL water three times (4500 rpm, 20 min). The final product was redispersed in 5 mL water for storage. The nanorods used for Raman experiments were obtained by taking 1 mL of storage solution and concentrating down to 100 μL.

***Synthesis of SDS-stabilized gold nanorods (zeta potentials: 16, 0 and -9 mV)***: To prepare SDS-stabilized gold nanorods with zeta potential of 16 mV, a sample of 1 mg of SDS was first added to a 50 mL centrifuge tube. Then, 5 mL storage solution of CTAB-stabilized gold nanorods was quickly added into the system at room temperature with vigorously shaking. The ligand exchange reaction happened instantly, and there were no obvious color changes of the solution during this time. The aliquots were further distributed into five vials, and the nanorods in each vial were cleaned by water at 6000 rpm for 20 min before concentrating down to 100 μL for Raman experiments. The SDS-stabilized gold nanorods with zeta potential of 0 or -9 mV can be synthesized using similar procedures by adjusting the SDS amounts to be 1-5 mg and reaction time to be 0-2 hours.

***Bacteria Cell Culture and Maintenance***: Cells were prepared as previously described.[2] Briefly, both types were streaked onto Trypticase™ Soy Agar with 5% Sheep Blood plates (BD #221239) from frozen stocks and incubated at 37°C without shaking for 24 hours. Individual colonies were picked and dispersed in 10 mL Lysogeny broth (LB) culture medium in 15 mL vials, followed by 15-hour incubation at 37°C with shaking at 300 rpm. Upon the incubation, 1 mL of culture was subject to 3 min washing with DI water for three times at 6000 rpm.

***Liquid Sample Well Fabrication***: The liquid well for SERS was fabricated following literature protocol with slight modifications. The silicon substrate was first cut from 4-inch silicon wafers to give 1 × 1 cm$^2$ dimension. Then, acetone and water were used to remove surface contaminants on silicon surface. 4 layers of 3M$^{TM}$ double-sized tape punched with Bostitch EZ Squeeze$^{TM}$ One-Hole Punch were further pasted onto the cleaned silicon substrate and cut into 1 × 1 cm$^2$ dimension. For bacteria-nanorod mixed samples, washed bacterial cells were mixed with gold nanorods at 1:1 ratio (5 μL each) to plate a total volume of 10 μL in the hole of liquid well. For pure bacteria/nanorod, the samples were prepared by directly mixing 3 μL of bacteria cells/nanorods with 3 μL of water. A glass coverslip (2.2 × 2.2 cm$^2$) was covered on top of all samples to avoid liquid evaporation.

***UV-Vis-NIR Absorption Spectroscopy***: The absorption spectra for all types of gold nanorods and bacteria were recorded on a Cary 6000 UV-Vis-NIR spectrophotometer at room temperature in the 300 – 1000 nm range. The quartz cuvette used for all measurements had 1.0 cm optical path length. Water was used as blank for background subtractions prior to all measurements.

***Raman Spectroscopy***: A commercial Raman microscope (WITec Alpha300) equipped with 785 nm laser was used for Raman measurement. Each liquid sample well filled with cells, gold nanorods or cell-gold nanorod mixtures only was illuminated for 120 sec at 40 mW with a single accumulation. Raman spectra were acquired using a 10X objective (Carl Zeiss; EC Epiplan-Neofluar DIC M27, 10X, NA = 0.90). The collected signals were recorded with a spectrometer (UHTS300 VIS-NIR spectrometer, WITec) equipped with an electron multiplying CCD (EMCCD, Andor DU970N-BV). Collected Raman spectra were subject to pre-processing,[3] including cosmic ray removal and denoising followed by baseline correction.[4]

***Transmission Electron Microscopy (TEM)***: TEM images were acquired on a FEI Tecnai (G2 Spirit TWIN) Digital TEM instrument at the MIT Material Research Laboratory. The samples were prepared by depositing gold nanorod solution via pipette onto a carbon film-coated 200-mesh copper grid (Ted Pella, Inc.) and further allowed to dry before measurements.

***Cryogenic Electron Microscopy (Cryo-EM)***: In sample preparation of 29 mV nanorod mixed with *E. coli*, 3 μL of the sample in buffer solution was applied to 200 mesh lacey carbon grid (LC200-Cu, Electron Microscopy Sciences, Hatfield, PA, USA). The grids were pre-treated with oxygen plasma using a Solarus 950 Gatan Advanced Plasma System. Excess sample on the grid was gently blotted using the Gatan Cryo Plunge III, followed by rapid plunging into liquid ethane to vitrify the sample. The grid was then mounted on a Gatan 626 single tilt cryo-holder, which was subsequently inserted into the TEM column. Both the specimen and the holder tip were maintained at cryogenic temperatures with liquid nitrogen to ensure preservation throughout the transfer and imaging process. Imaging was performed on a JEOL 2100 FEG microscope using a minimum dose method to reduce electron beam damage to the sample. The microscope was operated at 200 kV, with magnifications between 10,000x and 60,000x to evaluate particle size and distribution. All images were captured using a Gatan 2k x 2k UltraScan CCD camera.

In sample preparation of all other types of cell-nanorod mixtures, 3 µL of freshly prepared bacteria (diluted 1000 times) mixed with gold nanorods (bacteria: nanorods, 1:3, v/v) were applied to a 200 mesh lacey carbon grid (LC200-Cu, Electron Microscopy Sciences, Hatfield, PA, USA) that had been glow discharged for 60 seconds at 25 mA in a Quorum Emitech K100X glow discharger (Quorum Technologies, Judges House, UK) using a Vitrobot Mark IV (Thermo Fisher Scientific, Waltham, MA, USA). Grids were blotted with Vitrobot filter paper on the back side of the grid and a Vitrobot anti-contamination membrane (SubAngstrom, Brooklyn, NY, USA) facing the sample application side. Data was collected on a 200 kV Talos Arctica (Thermo Fisher Scientific, Waltham, MA, USA) at the Cryo-EM Facility at MIT.nano with a Falcon 3EC direct electron detector (Thermo Fisher Scientific, Waltham, MA, USA).

***Negative-Stain Electron Microscopy***: For negative-stain electron microscopy, 10 µL of the sample in buffer solution was applied to a 200-mesh copper grid coated with a continuous carbon film. After 60 seconds, excess solution was removed by gently touching the grid edge with a Kimwipe. Next, 10 µL of 1% aqueous phosphotungstic acid (negative stain) was applied to the grid. After 30 seconds, the excess stain was similarly removed, and the grid was air-dried at room temperature. The prepared grid was mounted on a JEOL single-tilt holder and inserted into the TEM column. The specimen was cooled with liquid nitrogen, and imaging was performed on a JEOL 2100 FEG microscope using a minimum dose method to minimize electron beam-induced damage. The microscope was operated at 200 kV, with magnifications ranging from 10,000x to 60,000x to assess particle size and distribution. All images were captured using a Gatan 2k x 2k UltraScan CCD camera.

***Zeta Potential Measurements***: Zeta Potential experiments for all types of gold nanorods and bacteria were conducted on a Malvern Zetasizer at room temperature. The samples were dispersed in water and further transferred to disposable folded capillary cells for measurements (Malvern, DTS 1070). At least three biological repeats were performed for each bacteria sample and the final zeta potential values were estimated based on the average of all measurements and standard deviations.

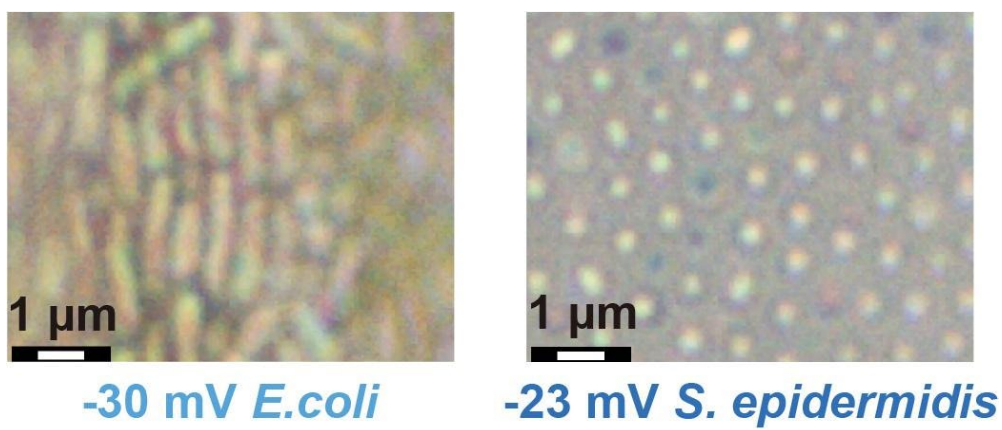

**Figure S1.** Bright Field microscopy images of rod-shaped *E. coli* (left) and round *S. epidermidis* (right) used for liquid SERS measurement in this study. The lengths of *E. coli* and *S.epidermidis* are approximately 2000 and 500 nm respectively.

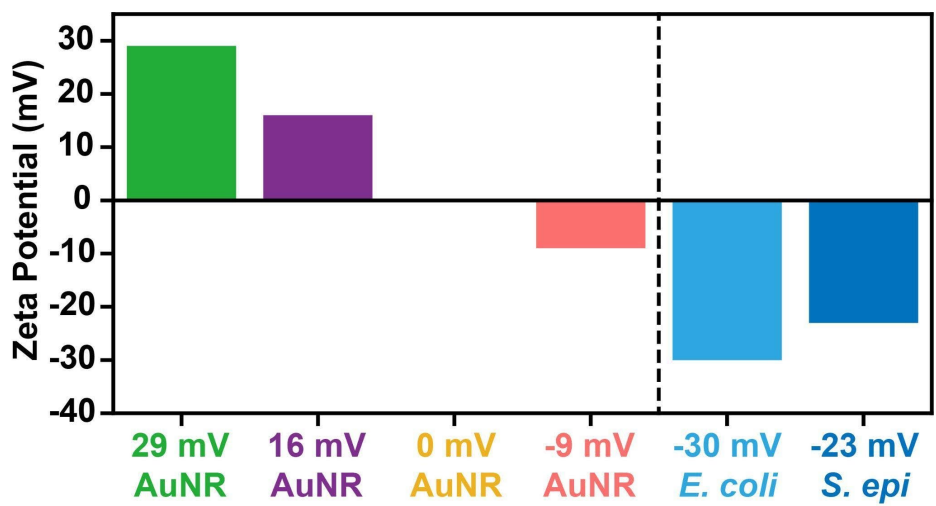

**Figure S2.** Zeta potential values of all types of gold nanorods and bacteria used in this study. At least three biological repeats were performed for zeta potential measurements of *E. coli* and *S. epidermidis*.

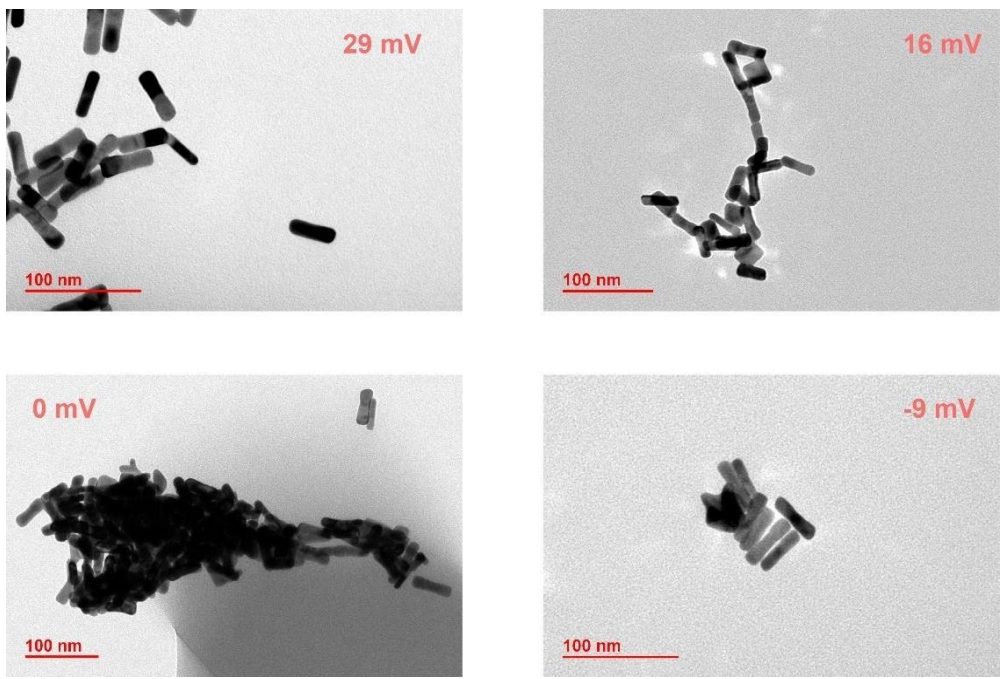

**Figure S3.** Large scale TEM images of the gold nanorods with zeta potentials of 29, 16, 0 and -9 mV. The 29 mV gold nanorods show the best stability in water. Slight coagulation was observed for both 16 and -9 mV gold nanorods. The 0 mV gold nanorods show severe coagulation due to low monodispersity.

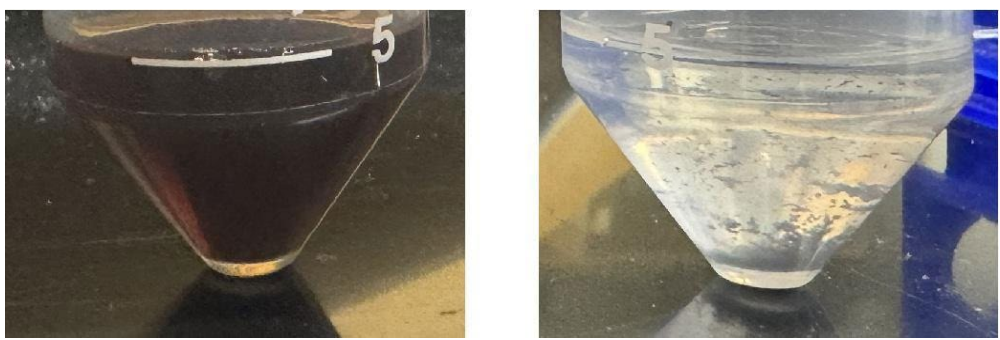

**Figure S4.** Photos of gold nanorods before (left, 29 mV) and after (right, 0 mV) ligands exchange of CTAB with SDS, showing the low stability of gold nanorods with zeta potential of 0 mV.

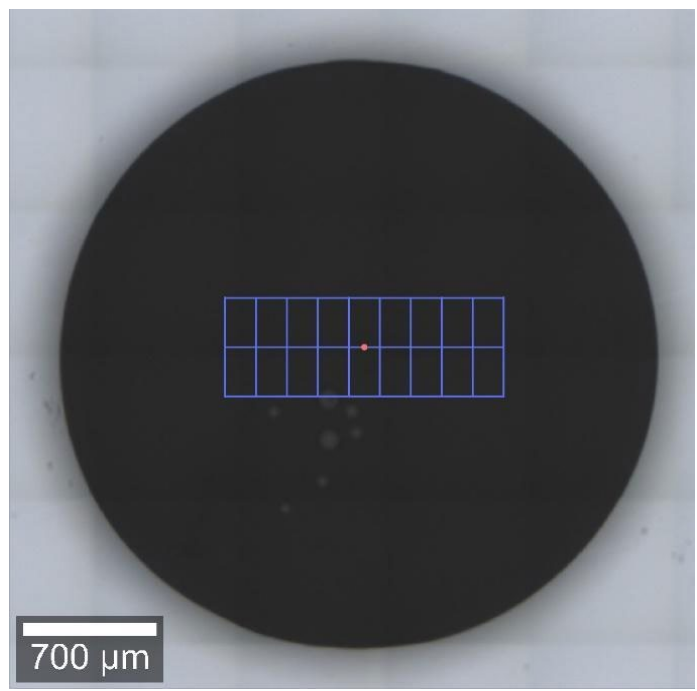

**Figure S5.** The Raman scanning area on the liquid well used in this study. The black circle is the liquid droplet of sample and the blue *xy* map region has a dimension of 2000 × 1000 μm².

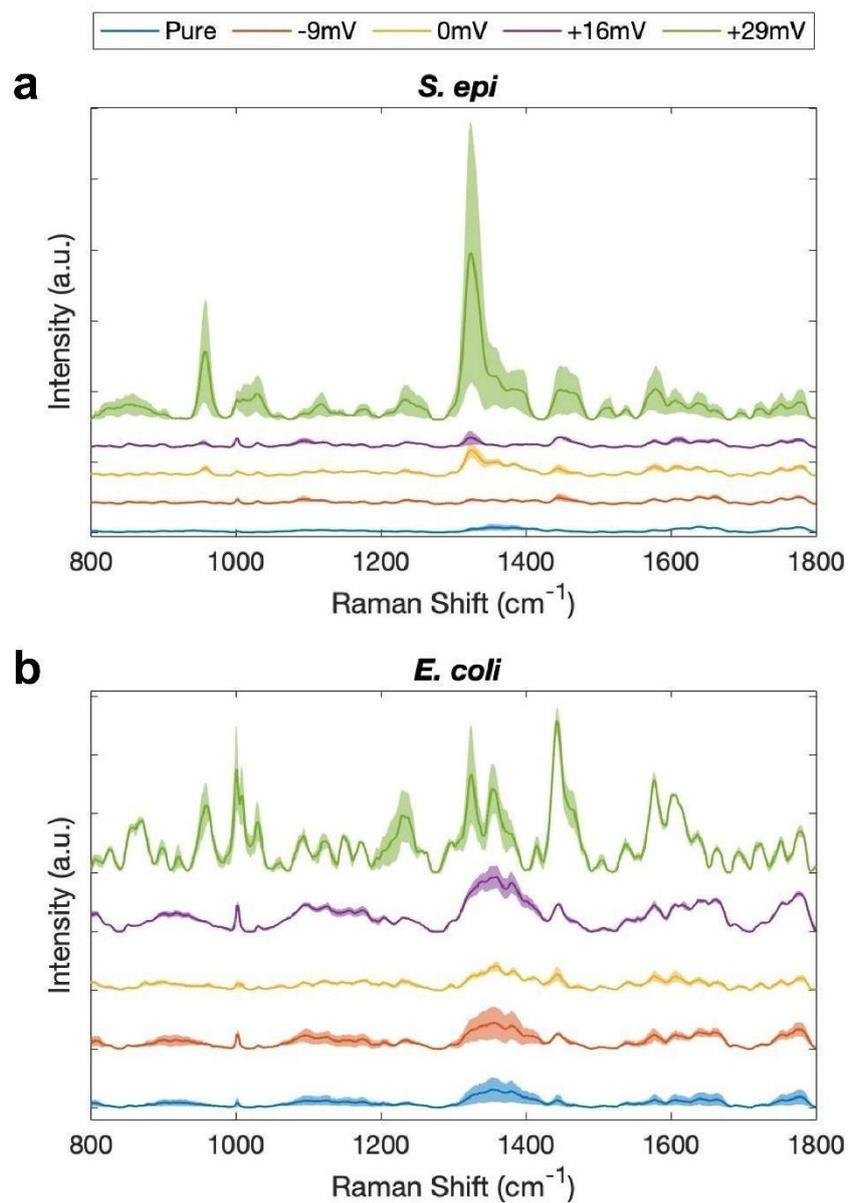

**Figure S6.** Additional Raman experiment repeats for pure (a) *S. epidermidis* and (b) *E. coli* cells as well as their bacteria-gold nanorod mixtures after background subtraction. Each line was averaged three additional biological repeats with 90 spectral data points.

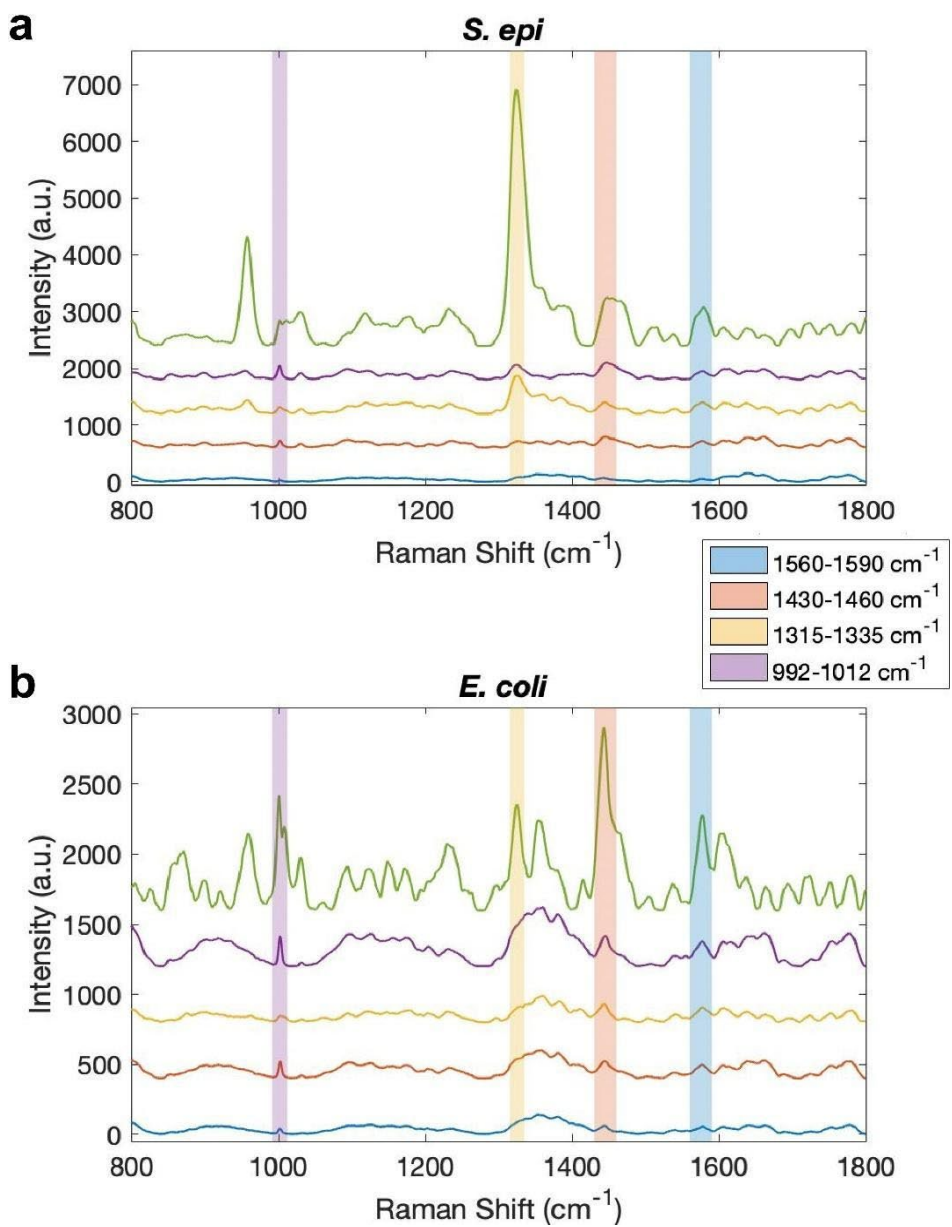

**Figure S7.** Additional Raman experiment repeats for pure (a) *S. epidermidis* and (b) *E. coli* cells as well as their bacteria-gold nanorod mixtures without background subtraction. The characteristic Raman bands were highlighted. Each line was averaged from three additional biological repeats with 90 spectral data points.

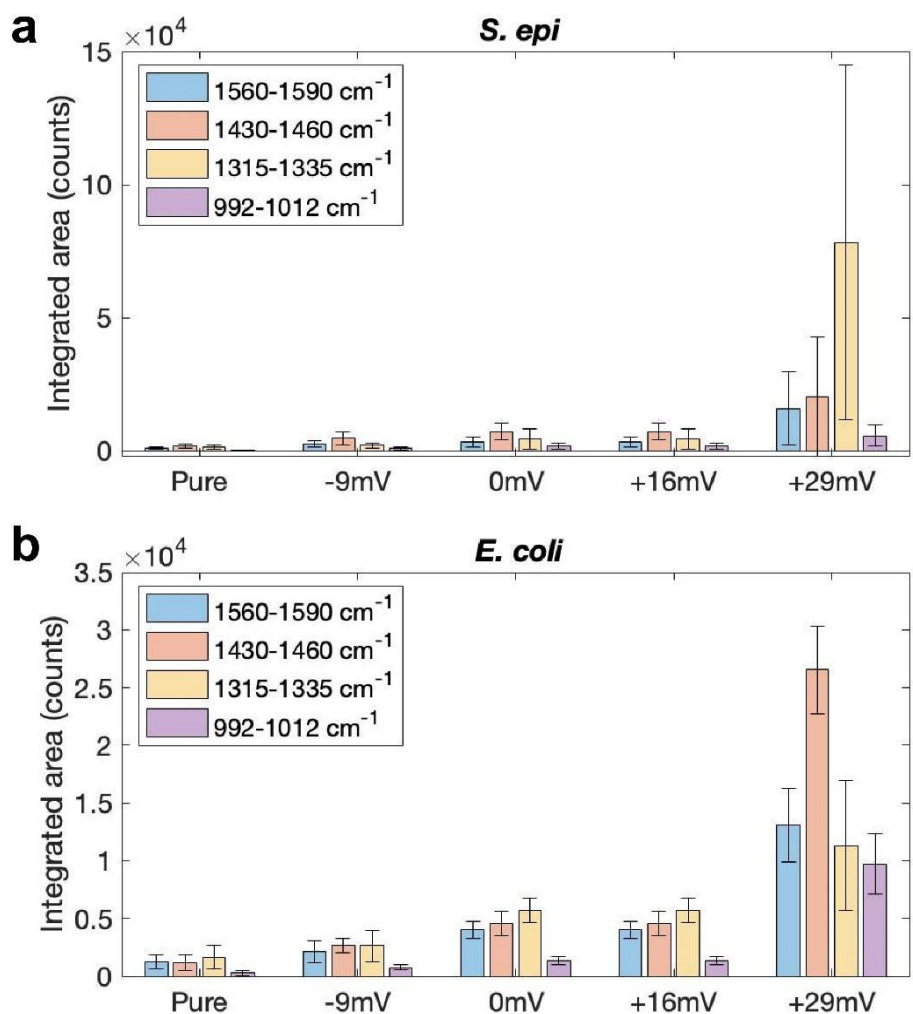

**Figure S8.** Areas under curves for characteristic Raman bands near 1000 cm$^{-1}$, 1350 cm$^{-1}$, 1450 cm$^{-1}$ and 1600 cm$^{-1}$, showing an increase trend of signal enhancement when the surface charge of gold nanorods goes from -9 to 29 mV. The bar plots shown here are for (a) *S. epidermidis* and (b) *E. coli* interacted with gold nanorods, respectively.

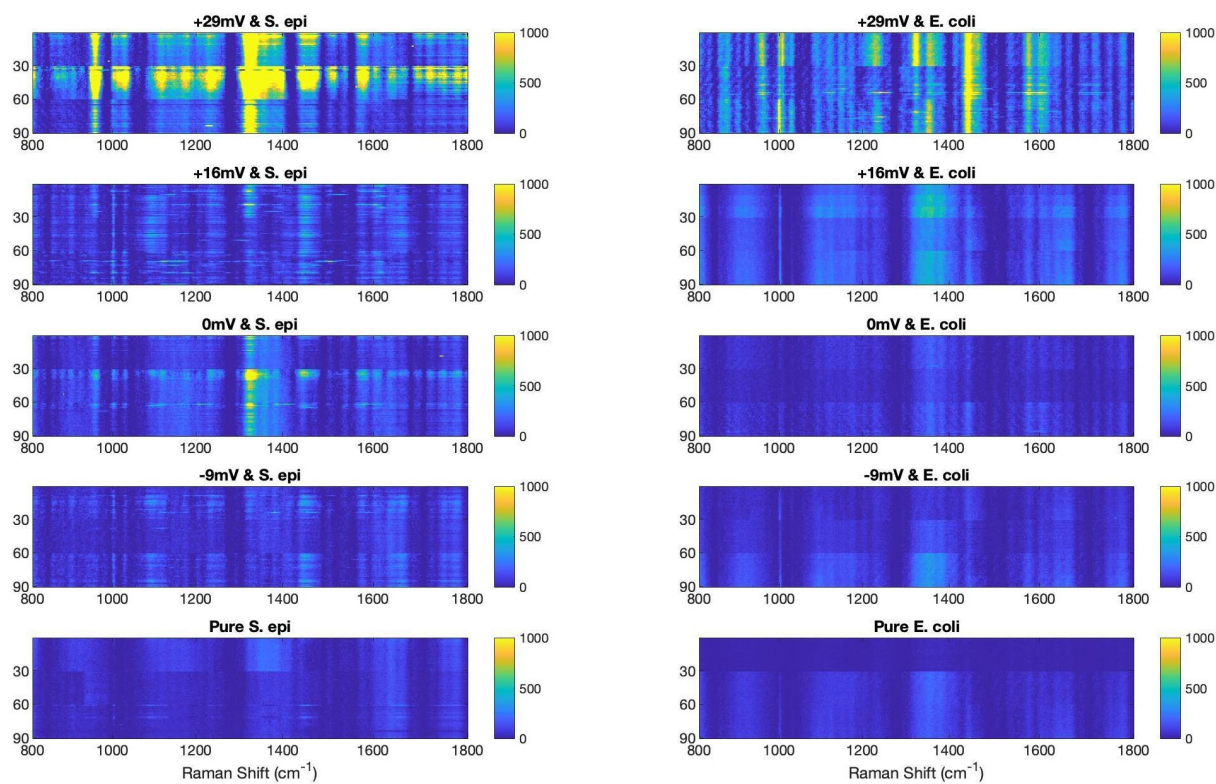

**Figure S9.** Additional heatmap plot spectra of *S. epidermidis* and *E. coli* mixed with gold nanorods with zeta potentials of 29, 16, 0 and -9 mV. Each bacteria-gold nanorod combination consists of 90 spectral data points from three additional biological repeats. The yellow or bright regions show higher signal intensity, indicating the enhancements for all combinations are consistent and gradually decrease as the surface charge drops.

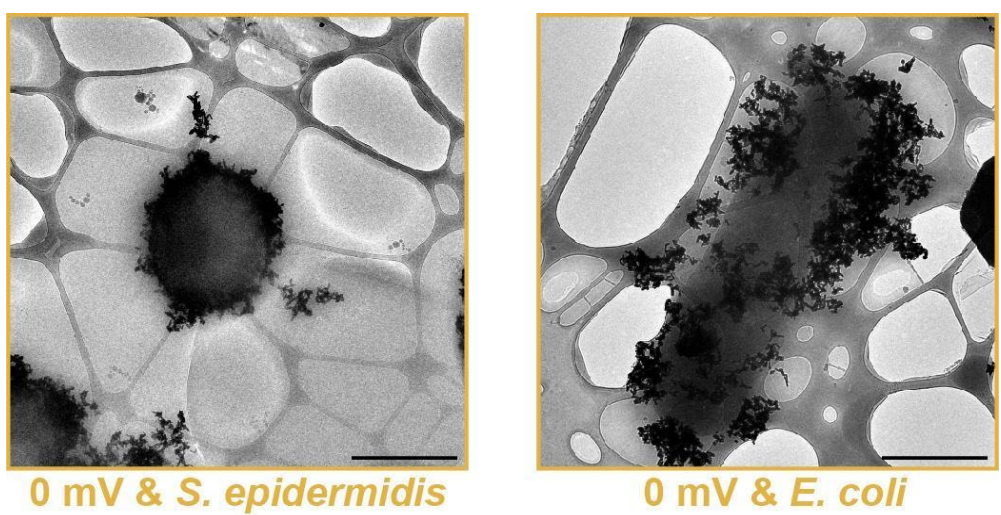

**Figure S10.** Cryo-EM images of 0 mV gold nanorods interacted with *S. epidermidis* (left) and *E. coli* (right) cells. Since there is no interparticle repulsion between 0 mV gold nanorods, they prefer to aggregate with each other and occasionally the cells may be wrapped inside.

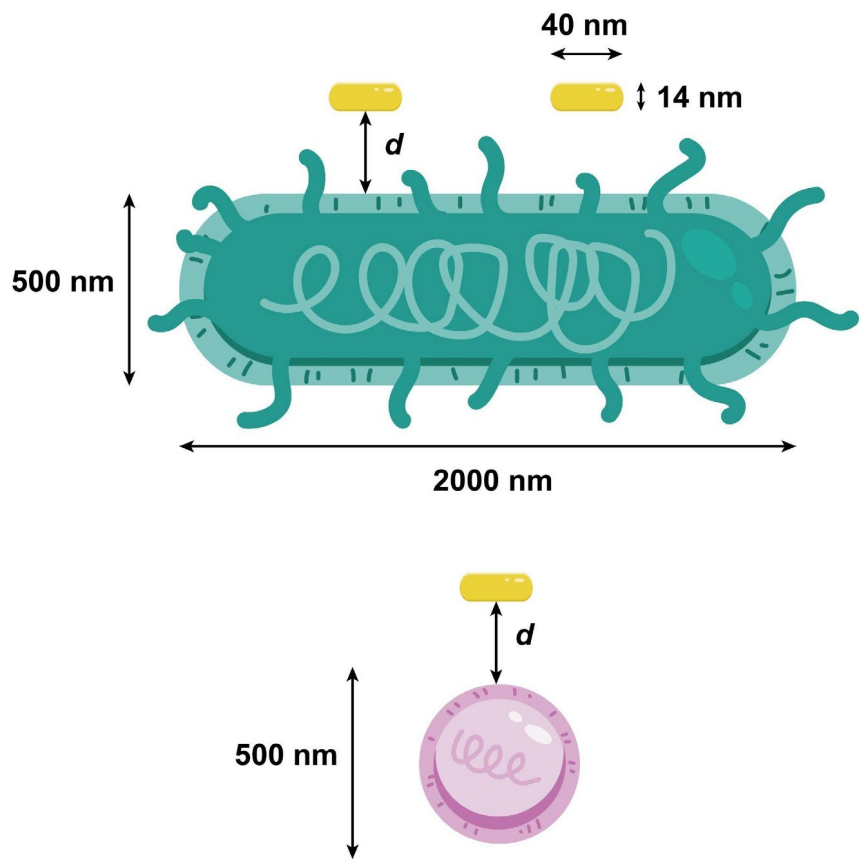

**Figure S11.** Sizes of *E. coli* (green), *S. epidermidis* (pink) and gold nanorods (yellow) used for the theoretical simulation in this study. The values were estimated based on the optical images of cells (Figure S1) and TEM images of gold nanorods (Figure 1c and S3). d is the separation distance between cell and gold nanorod surface.

**Table 1.** Code used for theoretical simulation in this study.

```
clear variables
close all

I = 1e-7;         % mol/L, ionic strength
inv_kappa = 0.304e-9/sqrt(I);
kappa = 1/inv_kappa;
% kappa = 1/5e-9;    % m, inverse debye length
epsilon0 = 8.854e-12;   % F/m, permittivity of free space
epsilon = 78*epsilon0;  % F/m, permittivity of water
NA = 6.023e23;         % Avogadro's constant
```

```matlab
kB = 1.38e-23;        % J/K, Boltzmann's constant
T = 273.15+25;        % K, Temperature
kT = kB*T;
% zetap = 29e-3;      % V, surface potential of particle
% zetab = -30e-3;     % V, surface potential of E.Coli
% zetab = -19e-3;     % V, surface potential of S.Epi
ap = 15e-9;
% ab = 500e-9;
% H = 1.72e-20;       % N.m, Hamaker constant
H = 3e-20;            % N.m, Hamaker constant
% H = 0;              % N.m, Hamaker constant
d = linspace(0.5,1000,3000)*1e-9;  % m, distances between particles
% d = linspace(0.5,50,1000)*1e-9;  % m, distances between particles
% d = linspace(0.2,20,1000)*1e-9;  % m, distances between particles
delGAB = 10e-3;       % J/m^2, acid-base interaction
lambda = 0.6e-9;      % m, correlation length of molecules
d0 = 0.157e-9;        % m, length of closest approach
Lambda = 3e14;

% % omega = 5.2e-7;   % kg/m^2, polymer density per unit area
% omega = 3e-6;       % kg/m^2, polymer density per unit area
% v2 = 0.91e-3;       % m^3/kg, polymer specific volume
% V1 = 107e-6;        % m^3/mol, solvent specific volume
% Theta = 160;        % Kelvin, theta-temperature
% psi1 = 0.3;         % Entropy of dilution parameter
% L = 30e-9;          % m, polymer contour length
Le = 200e-9;
Ls = 10e-9;

colours = parula(5);

names = ["E.Coli, 29mV",...
    "S.Epi, 29mV",...
    "E.Coli, 16mV",...
```

```
    "S.Epi, 16mV",...
    "E.Coli, 0mV",...
    "S.Epi, 0mV",...
    "E.Coli, -9mV",...
    "S.Epi, -9mV"];

zetab_vals = [-30,-19,-30,-19,-30,-19,-30,-19]*1e-3;
zetap_vals = [29,29,16,16,0,0,-9,-9]*1e-3;
omega_vals = [2.4,2.5,2.4,2.5,2.4,2.5,2.4,2.5]*0.22e-6;
L_vals =    [Le,Ls,Le,Ls,Le,Ls,Le,Ls];
ab_vals =   [650,500,650,500,650,500,650,500]*1e-9;

f = figure(1);
f.Position = [230,104,767,628];
hold on
xlabel('$d$ [nm]')
ylabel('$U$ [$k_\mathrm{B} T$]')
legend
xlim([0,250]);
ylim([-120,20]);

f = figure(2);
f.Position = [230,104,767,628];
hold on
xlabel('$d$ [nm]')
ylabel('$U$ [$k_\mathrm{B} T$]')
legend
xlim([0,20]);
ylim([-120,10]);

f = figure(3);
f.Position = [230,104,767,628];
hold on
xlabel('$d$ [nm]')
```

```matlab
ylabel('$U$ [$k_\mathrm{B} T$]')
legend

xlim([0,250]);
ylim([-70,20]);

figure(4);
hold on
tiledlayout('flow');

f = figure(5);
f.Position = [230,104,767,628];
hold on
xlabel('$\psi$ [mV]')
ylabel('$d_\mathrm{min}$ [nm]')

secondmin = zeros(size(ab_vals));
for ii = 1:length(names)
% for ii = 1:2:length(names)
% for ii = 4

%     figure();
%     hold on

    ab = ab_vals(ii);
    zetab = zetab_vals(ii);
    zetap = zetap_vals(ii);
    omega = omega_vals(ii);
    L = L_vals(ii);

    VLW = -H*(ap*ab)./(6*d*(ap+ab));
    VEL = pi*epsilon*ap*ab*(zetap^2+zetab^2)/(ap+ab) ...
        *((2*zetap*zetab)/(zetap^2+zetab^2)* ...
        log((1+exp(-kappa*d))./(1-exp(-kappa*d))) ...
```

```matlab
            + log(1-exp(-2*kappa*d)));
        VAB = 2*pi*ap*ab/(ap+ab)*lambda*delGAB*exp((d0-d)/lambda);
%       VAB = 0;
%       S = 2*log(L./d) + 2*d/L - 2;
%       VS = 2*pi*(ap*ab)/(ap+ab)*omega^2*NA*(v2^2/V1)*psi1*(1-Theta/T)*S*kT;
%       VS(d>L) = 0;
        VS = 50*(ap*ab)/(ap+ab)*Lambda^(3/2)*L^2*exp(-2*pi*d/L)*kT;
%       VS = 0;

        VT = (VLW+VEL+VAB+VS)/kT;

        mins = find(islocalmin(VT));
        secondmin(ii) = d(mins(end));

        if mod(ii,2)==0
            symbol = '-';
        else
            symbol = ':';
        end

        figure(1);
        hold on
        plot(d*1e9, VT, symbol,...
            'DisplayName',names(ii),...
            'Color',colours(ceil(ii/2),:));

        if mod(ii,2)==0
            figure(2);
            hold on
            plot(d*1e9, VT, symbol,...
                'DisplayName',names(ii),...
                'Color',colours(ceil(ii/2),:));
        end
```

```matlab
    if mod(ii,2)==1
        figure(3);
        hold on
        plot(d*1e9, VT, symbol,...
            'DisplayName',names(ii),...
            'Color',colours(ceil(ii/2),:));
    end

    if ii==1
        HV = 'on';
    else
        HV = 'off';
    end

    figure(4);
    hold on
    nexttile
    hold on

    plot(d*1e9, VT, symbol,...
        'DisplayName',names(ii),...
        'Color',colours(ceil(ii/2),:));

    plot(d*1e9, (VLW)/kT, 'k:', 'DisplayName','Van der Waals',...
        'HandleVisibility',HV);
    plot(d*1e9, (VEL)/kT, 'k--', 'DisplayName','Electrostatic',...
        'HandleVisibility',HV);
    plot(d*1e9, (VAB)/kT, 'k-.', 'DisplayName','Acid-base',...
        'HandleVisibility',HV);
    plot(d*1e9, (VS)/kT, 'k-', 'DisplayName','Steric',...
        'HandleVisibility',HV);
    xlabel('$d$ [nm]')
    ylabel('$U$ [$k_\mathrm{B} T$]')
```

```matlab
    legend
%     ylim([-15,5]);

    figure(5);
    hold on
    if mod(ii,2)==0
        plot(zetap*1e3, secondmin(ii)*1e9, 'o', ...
            'Color',colours(ceil(ii/2),:), ...
            'MarkerFaceColor',colours(ceil(ii/2),:),...
            'MarkerSize',10, 'DisplayName',names(ii));
    else
        plot(zetap*1e3, secondmin(ii)*1e9, 's', ...
            'Color',colours(ceil(ii/2),:), ...
            'MarkerFaceColor',colours(ceil(ii/2),:),...
            'MarkerSize',10, 'DisplayName',names(ii));
    end

end

% figure(1);
% annotation('textbox', 'String', ...
%     {sprintf('$1/\\kappa = %0.4g$ [nm]', inv_kappa*1e9), ...
%      sprintf('$a_\\mathrm{particle} = %0.4g$ [nm]', ap*1e9), ...
%      sprintf('$H = %0.4g$ [N.m]', H), ...
%      sprintf('$L_\\mathrm{polymer} = %0.4g$ [nm]', L*1e9), ...
%      sprintf('$\\Delta G_{AB} = %0.4g$ [J/m$^2$]', delGAB)});

% for pp = 1:length(Phibp_vals)
%     Phip = Phibp_vals(pp);
%     Q = DLVO_potential(d, epsilon, Phip, Phib, H, kappa);
%     U = 2*pi*Q*B;
%
%     dscaled = d*1e9;
```

```
%     Uscaled = U/4e-21;
%
%     plot(dscaled,Uscaled, '-',...
%         'Color',colours(pp,:),...
%         'LineWidth',3,...
%         'DisplayName',names(pp));
%
%     TF = islocalmin(Uscaled);
%     scatter(dscaled(TF), Uscaled(TF), 500,colours(pp,:), 'p', 'filled',...
%         'HandleVisibility','off');
% end

% ylim([-5,5]);

% legend
%
% figure();
% hold on
%
% xlabel('')

% legend('Num')

%% surface charge densities
sigP = -kappa*epsilon*zetap_vals;
sigB = -kappa*epsilon*zetab_vals;

1.60217663e-19;

% %% self-interactions
%
% figure();
% hold on
```

```matlab
%
% for ii = 1:length(zetap_vals)
%     zetab = zetab_vals(ii);
%     zetap = zetap_vals(ii);
%     omega = omega_vals(ii);
%     L = L_vals(ii);
%
%     VLW = -H*(ap*ap)./(6*d*(ap+ap));
%     VEL = pi*epsilon*ap*ap*(zetap^2+zetap^2)/(ap+ap) ...
%         *((2*zetap*zetap)/(zetap^2+zetap^2)* ...
%         log((1+exp(-kappa*d))./(1-exp(-kappa*d))) ...
%         + log(1-exp(-2*kappa*d)));
%
%     plot(d*1e9, (VLW+VEL)/kT, symbol,...
%         'DisplayName',names(ii),...
%         'Color',colours(ii,:));
%
%
% end
%
% xlabel('$d$ [nm]')
% ylabel('$U$ [$k_\mathrm{B} T$]')

% %%% functions
% function Q = DLVO_potential(d, epsilon, Phi1, Phi2, H, kappa)
%
%     Q = 2*epsilon*Phi1*Phi2.*exp(-d*kappa) - H./(12*pi*d);
%
% end
```

**Table S2.** Values of minimum separation distance and corresponding interaction energy for different cell-nanorod mixtures shown in Figure 4b and c.

| Zeta potential of nanorods (mV) | Bacteria type | Separation distance (nm) | Interaction energy ($k_BT$) |
|---|---|---|---|
| +29 | E. coli | 89.2 | -56.2 |
| | S. epidermidis | 1.8 | -111.6 |
| +16 | E. coli | 113.1 | -30.8 |
| | S. epidermidis | 2.2 | -60.7 |
| 0 | E. coli | 152.8 | -7.9 |
| | S. epidermidis | 2.5 | -18.9 |
| -9 | E. coli | 193.8 | 1.5 |
| | S. epidermidis | 2.8 | -5.0 |

**Table S3.** Values of minimum separation distance and corresponding interaction energy for different cell-nanorod mixtures shown in Figure 6c and d.

| Zeta potential of nanorods (mV) | Bacteria type | Separation distance (nm) | Interaction energy ($k_BT$) |
|---|---|---|---|
| +29 | E. coli | 89.2 | -56.2 |
| | S. epidermidis | 109.1 | -34.2 |
| +16 | E. coli | 113.1 | -30.8 |
| | S. epidermidis | 136.5 | -17.0 |
| 0 | E. coli | 152.8 | -7.9 |
| | S. epidermidis | 189.1 | -2.8 |
| -9 | E. coli | 193.8 | 1.5 |
| | S. epidermidis | | |

**Table S4** The zeta potential of bacteria and gold nanorods and their corresponding sum values shown in Figure 7.

| Zeta potential of E. coli (mV) | Zeta potential of S. epidermidis (mV) | Zeta potential of gold nanorods (mV) | Sum Value (mV) |
|---|---|---|---|
| -30 | | 29 | -1 |
| -30 | | 16 | -14 |
| -30 | | 0 | -30 |
| -30 | | -9 | -39 |
| | -23 | 29 | 6 |
| | -23 | 16 | -7 |
| | -23 | 0 | -23 |
| | -23 | -9 | -32 |